\documentclass[doublecol]{epl2} 

\usepackage{latexsym}

\usepackage{graphicx}
\usepackage{amssymb}
\usepackage{epstopdf}
\usepackage{float}
\title{Capillary-Gravity Waves on Depth-Dependent Currents:\\
Consequences for the Wave Resistance}
\shorttitle{Capillary-Gravity Waves on Depth-Dependent Currents}

\author{M. Benzaquen\inst{1} \and E. Rapha\"el\inst{1}}
\shortauthor{M. Benzaquen \etal}

\institute{                    
  \inst{1} Laboratoire PCT - UMR  Gulliver CNRS 7083, ESPCI, 10 rue Vauquelin, 75005 Paris, France
}

\pacs{47.35.-i}{Hydrodynamic waves}
\pacs{68.03.-g}{Gas-liquid and vacuum-liquid interfaces }

\abstract{ We study theoretically the capillary-gravity waves created at the water-air interface by a small two-dimensional perturbation 
in the frequently encountered case where a depth-dependent current is present in the fluid.
Assuming linear wave theory, we derive a general expression of the wave resistance experienced by the perturbation as a function of the current profile in the case of an inviscid fluid. We then illustrate the use of this expression in the case of constant vorticity.}

\begin{document}

\maketitle

%%%%%%%%%%%%%%%%%%%%%%%%%%%%%%%%%%%%%%%%%% INTRO

Water waves are both fascinating and of great practical importance \cite{Lighthill,Lamb, Johnson}. 
For these reasons, they have attracted the attention of scientists for many centuries \cite{Darrigol}. 
Water waves can for instance be generated by the wind at sea, by a moving boat on a calm lake, 
or simply by throwing a pebble into a pond. 
Their propagation at the surface of water is driven by a balance between 
the liquid inertia and its tendency, under the action of gravity or of surface tension 
(or a combination of both in the case of {\it{capillary-gravity waves}}),
 to return to a state of stable equilibrium \cite{LandauLifshitz}. 
 Neglecting the viscosity of water, the dispersion relation of linear 
capillary-gravity waves  relating the angular frequency $\omega$ to the 
wavenumber $k$ is given by 
$\omega^{2} = \left (gk+\gamma k^{3}/\rho \right) \tanh (k h)$,
where $\gamma$ is the liquid-air 
surface tension, $\rho$ the liquid density, $g$ the acceleration due to gravity
and $h$ the depth of water \cite{Lighthill}.
The above equation may also be written as a dependence
of the phase velocity $c = \omega/k$ on the wavenumber:
$c(k) = {\left(g/k + \gamma k/\rho\right)}^{1/2} \left(\tanh(k h)\right)^{1/2} $.
The dispersive nature of capillary-gravity waves is
responsible for the complicated wave pattern generated at the free
surface of a still liquid by
a moving disturbance such as a partially  immersed
object   ({\it{e.g.}} a boat or an insect) or 
an external surface pressure source. The propagating waves generated by the moving disturbance
continuously remove energy to infinity. Consequently,
the disturbance will experience a drag, ${R}$, called the
{\it wave resistance} \cite{Havelock}. In the case of boats and
large ships, this drag is known to be a major source of resistance
and important efforts have been devoted to the design of
hulls minimizing it \cite{Milgram}. The case of objects small relative
to the capillary length $\kappa^{-1} = {\left(\gamma/(\rho g)\right)}^{1/2}$
has only recently been considered \cite{Elie:96, Elie:99, Keller,
Chevy, Bacri, Steinberg1, Closa} and has attracted strong interest
in the context of insect locomotion on water surfaces \cite{Bush, Voise}.
In the case of a pressure distribution of amplitude $p$ localized along a line 
and traveling over the surface with speed $V$ perpendicularly 
to  its length, the wave resistance for deep water ($h \to +\infty$)
is given by ${R} = (p^2/\gamma) {\left(1 - (c_{\rm{min}}/V)^4 \right)}^{1/2}$
for $V >c_{\rm{min}}$, 
and ${R} = 0$ for $V < c_{\rm{min}}$ \cite{Lamb, Elie:96}.
Here  $c_{\rm{min}} = (4 g \gamma  /\rho)^{1/4}$ is the minimum of
the wave velocity $c(k)$  for deep-water capillary-gravity waves. In the limit 
$V \gg c_{\rm{min}}$, the wave resistance reduces to $p^2/\gamma$.
Note that as $V$ approaches $c_{\rm{min}}$ (from above), the wave 
resistance becomes unbounded. This diverging behavior of  the wave resistance
(in the case of a two-dimensional pressure distribution) is related to
the fact that when $V$ approaches $c_{\rm{min}}$, the phase velocity $\omega(k)/k$
(equal to $V$, see  \cite{Elie:96}) and the group velocity ${\rm{d}}\omega(k)/{\rm{d}}k$
 tend towards the same value. It was shown in  \cite{Elie:99} that in the presence
of viscosity, the wave resistance remains bounded as $V$ approaches $c_{\rm{min}}$.
A similar regularization also exists when one takes into account non-linear effects \cite{Dias}.

In many cases of physical interest (like wind-generated flows \cite{Kawai, Zeisel}), the waves
propagate on shear currents rather than in still water (see, {\it{e.g.}}, the seminal work
of Miles \cite{Miles}).
The general problem of the interaction between 
water waves and arbitrary steady current is of great physical significance \cite{Peregrine}.
It is, however, rather difficult and remains
largely unsolved \cite{Kirby, Valenzuela, Victor, Zhang, helena}.
In this letter, we consider how the above predictions for the wave resistance
are modified when the pressure distribution propagates on steady shear currents
 (taking also into account the effect of water finite depth, 
 assuming $h > \sqrt{3/2} \, \kappa^{-1}$). 
 
 The present letter is organized as follows. 
 We first formulate the problem by analyzing the fluid equation of motion together with
 the boundary conditions. This allows us to get the free surface displacement as a 
 function of the shear current and the pressure disturbance. We then establish a 
 general expression for the wave resistance experienced by the perturbation as a function of the current profile.
  Finally, we illustrate the use of this expression in the case of constant vorticity.

%%%%%%%%%%%%%%%%%%%%%%%%%%%%%%%%%%%%%%%%%% CALCUL
 \smallskip
 
\begin{figure}
\begin{center}
\includegraphics[width= 0.4 \columnwidth]{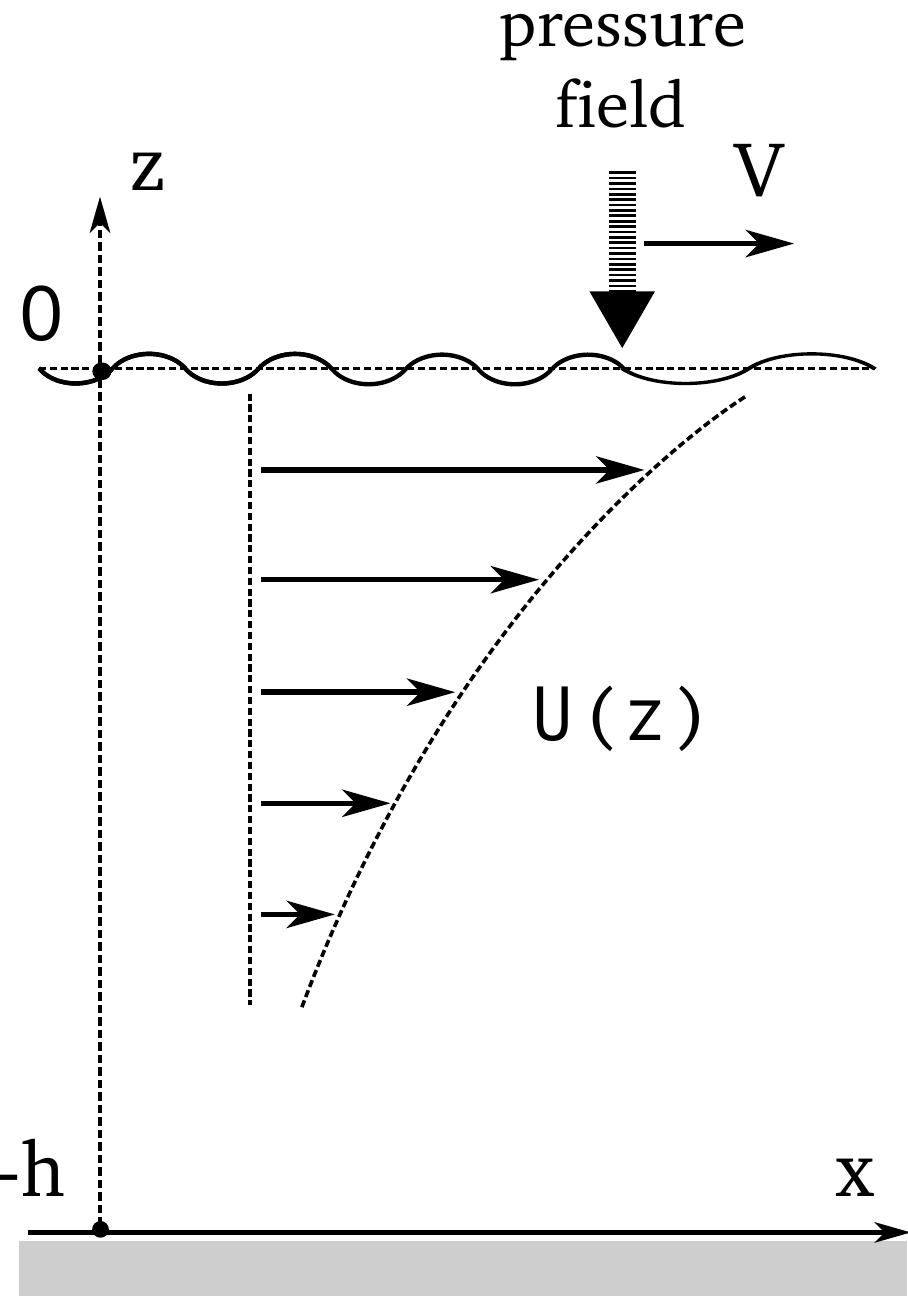}
\end{center}
\caption{Schematic diagram: an external surface pressure distribution moves with speed $V$ in the $x$ direction above the free surface of a liquid of depth $h$ in which a  velocity profile $U(z)$ preexists.} 
\label{fig1}
\end{figure}

{\it{Model - }}  We consider the {\it {two-dimensional}} motion of a layer of fluid (assuming the fluid
to be incompressible and inviscid).
This implies that all physical quantities depend spatially on only one horizontal coordinate denoted by $x$,
and on the vertical coordinate denoted by $z$. The flat bottom is given by $z = -h$
while the free surface (in the absence of waves, see below) corresponds to $z = 0$.
We assume the existence of a steady shear current below the free surface characterized
by a velocity component $U(z)$ in the horizontal $x$-direction, and a velocity component
equal to zero in the vertical  $z$-direction. In addition to this flow, capillary-gravity waves are generated
by a pressure distribution (invariant along the $y$-direction) moving with a constant 
speed $V$ along the $x$-direction, as indicated in Fig.\ref{fig1}. Let $z=\zeta(x,t)$ denote the displacement 
of the free surface (in the presence of waves). The  velocity component in the horizontal $x$-direction is now given by
$U(z) + u(x,z,t)$, and the velocity component in the vertical $z$-direction by $w(x,z,t)$.
The additional velocity field $(u, w)$ is assumed to be a
first order correction to the undisturbed flow. Associated with the wave-induced motion is a stream function $\psi(x,z,t)$ 
so that $u=\partial_z \psi$ and $w=-\partial_x \psi$. Note that the use of the stream function does 
not impose an irrotational fluid motion.

 Having in mind that the phase velocity  is imposed by
 the velocity $V$ of the pressure disturbance \cite{Lighthill,Elie:96},
we shall seek a stream function of the form
 \begin{eqnarray}
\psi(x,z,t) &=&\int \frac{dk}{2\pi} \hat A(k)\,f_k(z) \,e^{ik(x-Vt)}.
\label{psi}
 \end{eqnarray}
We shall now determine the function $f_k(z)$ (characterizing the $z$-dependence 
of the stream function) using the equation of motion and the boundary conditions.
According to Euler's equation we have
 \begin{eqnarray}
\rho(  \partial_t u +U(z)\partial_xu+wU'(z))&=&- \partial_x p,  \\ 
\rho(\partial_t w +U(z)\partial_xw)&=&- \partial_z p-g.
 \end{eqnarray} 
Using (\ref{psi}), the above two equations can be 
rewritten as
 \begin{eqnarray}
&\displaystyle  \int \frac{dk}{2\pi} \hat A(k)\,ik\rho\{ (V-U)f'_k+U'f_k\} \,e^{ik(x-Vt)}=\,\partial_x p,\,\,\,   \label{suivantx}\\
 &\displaystyle \int \frac{dk}{2\pi} \hat A(k)\,\rho k^2\{ (V-U) f_k\} \,e^{ik(x-Vt)}=\,\partial_z p+\rho g.
 \,\,\, \,\,\,  \label{suivantz}
 \end{eqnarray} 
Eliminating pressure between equations  (\ref{suivantx}) and (\ref{suivantz}) yields
 \begin{eqnarray}
\big(V - U(z)\big) \big[f''_k(z) - k^2f_k(z)\big] \, + \, U''(z) f_k(z) \, = \, 0.
\label{OS}
 \end{eqnarray} 
Equation (\ref{OS}), which relates the function $f_k(z)$ to the current profile $U(z)$, is known as the inviscid Orr-Sommerfeld or Rayleigh equation \cite{Kirby, helena}. It has 
to be supplemented with boundary conditions.
 At the bottom, the fluid velocity vanishes and so $f_k(z=-h)=0$. 
Let $U_0 \equiv U(z=0)$ and $U_0' \equiv \partial_zU(z=0)$. The dynamic free surface 
boundary condition \cite{Lamb} can be written as
 \begin{eqnarray}
& \displaystyle \int \frac{dk}{2\pi} \hat A(k)\,\rho\{ -(U_0-V)f'_k(0)+U_0'f_k(0)\} \,e^{ik(x-Vt)} \nonumber \\
& =\,\,p +\rho g \zeta,
\label{int2}
\end{eqnarray}
where, according to Laplace's formula, the pressure $p$ equals $p_{ext}(x,t)-\gamma\partial^2_x \zeta$ \cite{LandauLifshitz}. Note that $f_k(0)$ and $f_k'(0)$ depend on the full profile $U(z)$ with $z \in [-h,0]$ (see (\ref{OS})).
Let
$p_{ext}(x,t)={(2\pi)^{-1}}\int {dk} \, {\hat P_{ext}(k)} \,e^{ik(x-Vt)}$ and
$\zeta(x,t)={(2\pi)^{-1}}\int {dk} \, {\hat \zeta(k)} \,e^{ik(x-Vt)}$.
Using then the kinematic  free surface boundary condition $w=\partial_t \zeta +U_0 \partial_x \zeta$ we obtain
$\hat A(k) =-({U_0-V})\hat \zeta(k)/{f_k(0)}$ and 
\begin{eqnarray}
&\rho\left\{ (U_0-V)^2\displaystyle \frac{f'_k(0)}{f_k(0)}   -U'_0 (U_0-V)-\left( \displaystyle g+\frac{\gamma k^2}{\rho}  \right)  \right\} \hat \zeta(k)  \nonumber\\
&=\,{\hat P_{ext}(k)}.
\label{versPGG}
\end{eqnarray}
Equation (\ref{versPGG}) is of physical importance since for a given current profile it relates the Fourier component  ${\hat \zeta(k)}$ of the surface displacement to
the Fourier component ${\hat P_{ext}(k)}$ of the pressure disturbance.

Let us emphasize that despite the fact that we are working within the frame of a linear wave theory,  a linear combination of solutions corresponding to different current profiles will generally not be a solution of the problem at hand. See \cite{Kirby,helena} for further details on this issue.

\smallskip

%%%%%%%%%%%%%%%%%%%%%%%%%%%%%%%%%%%%%%%%% WR

{\it{Wave resistance - }} We can now investigate the wave resistance $R$ experienced by the disturbance. According to Havelock \cite{Havelock}, we may imagine a rigid cover fitting the surface everywhere. The pressure $p_{ext}(x)$ is applied to the liquid surface by means of this cover; hence the wave resistance is simply the total resolved pressure in the $x$ direction. This leads to $R=-\int dx \,\,p(x)  \,\partial_x \zeta (x)$ \cite{Havelock}.
According to (\ref{versPGG}) the wave resistance can then be written as

\medskip

$R = $
\begin{eqnarray}
\displaystyle  \int \frac{dk}{2\pi\rho} \frac{-ik  | \hat P_{ext}(k)|^2 }{ (U_0-V)^2\displaystyle  \frac{f'_k(0)}{f_k(0)}     -U'_0 (U_0 - V)- \left( g +\frac{\gamma k^2}{\rho} \right)},\,\,\,\label{WR}
 \end{eqnarray}
 %\end{widetext}
 
 which is the central result of the present letter. It allows one to calculate the wave resistance experienced by the moving disturbance for any current profile $U(z)$. Note that if one replaces $V$ by $\omega/k$ and sets the denominator of
 (\ref{WR}) to zero, one gets a general formula for the dispersion relation of capillary-gravity waves  on
 depth-dependent current.
 
\smallskip
 %%%%%%%%%%%%%%%%%%%%%%%%%%%%%%%%%%%%%%%%% SWITCHING

 \begin{figure}[H]
\begin{center}
\includegraphics[width= 1 \columnwidth]{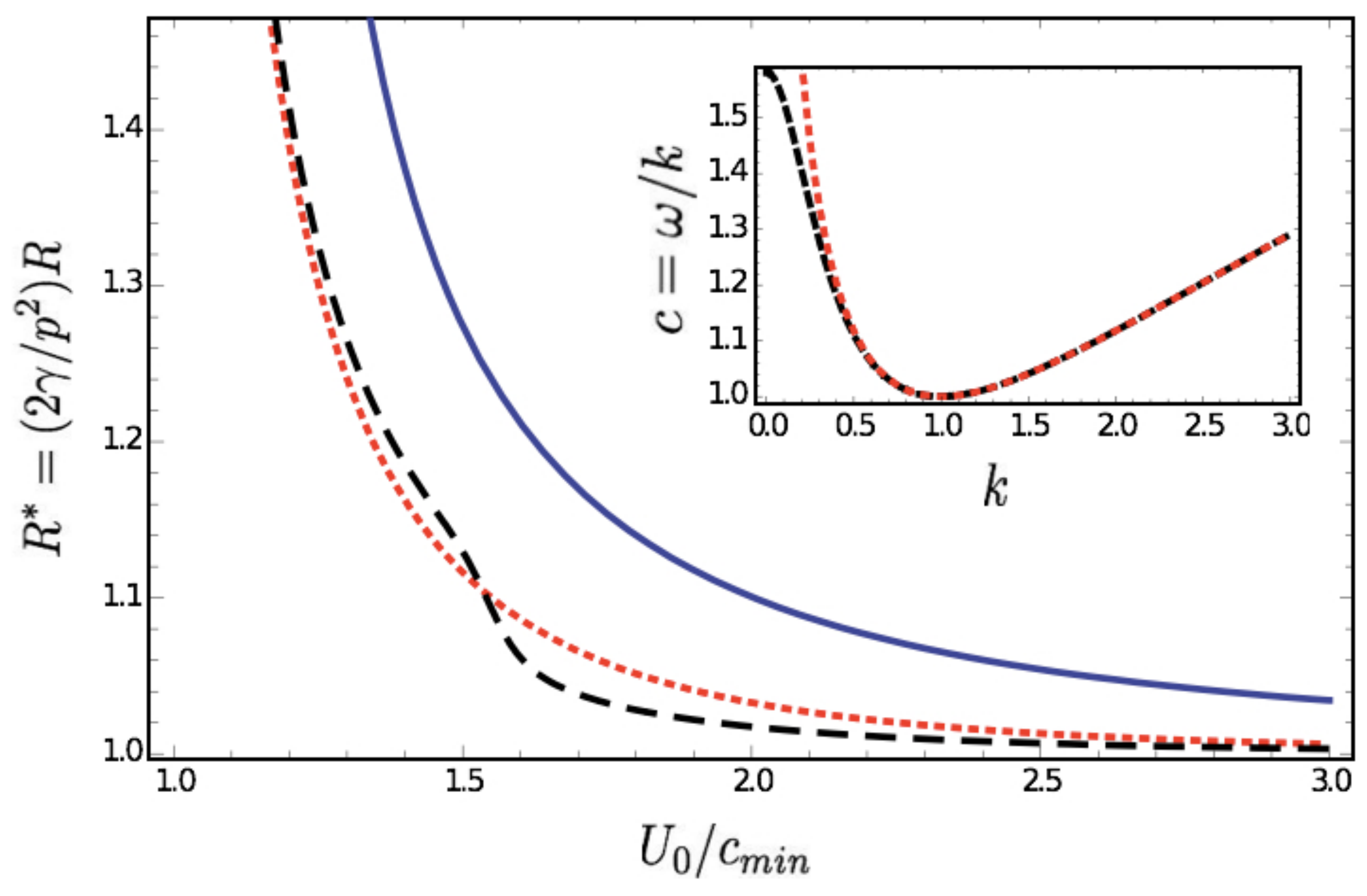}
\end{center}
\caption{(Color online) Plot of the wave resistance $R^*$ for $V=0$ in units
of $p^2 /2\gamma$, as a function of the reduced velocity $U_0/c_{\rm{min}}$. The red curve
(dotted) corresponds to a uniform current $U(z)=U_0$ with infinite depth, the black one (dashed) corresponds  also to a uniform current $U(z)=U_0$ with finite depth $h=5 \kappa^{-1}$, while the blue one (solid line)  corresponds to a linear current $U(z)=U_0(h+z)/h$ with the same finite depth $h=5 \kappa^{-1}$.  The insert shows the corresponding dispersion relations.}
 \label{fig2}
\end{figure}

 The integral in equation (\ref{WR}) cannot be evaluated unambiguously because the poles of the integrand are on the domain of integration.
 This ambiguity can be removed by imposing the \textit{radiation condition} that there be no wave coming in from infinity. There are several mathematical procedures equivalent to this radiation condition. One way is to consider that the amplitude of the disturbance has increased slowly to its present value in the interval $-\infty\leq t\leq0$ : $p_{ext}^{\epsilon}(x,t) =e^{\epsilon t} \,p_{ext}(x,t)$
where $\epsilon$ is a small positive number that will ultimately be allowed to tend to zero. 
%The wave resistance (\ref{WR}) is now implicitly replaced by $R=\lim_{\epsilon\rightarrow0} R^{\epsilon}$. 
The poles of the integrand have been shifted over and above the real axis. Since the poles are now out of the domain of integration, the integral can be evaluated numerically unambiguously.
 
For a non linear current profile, \textit{i.e.} $U''(z)\neq0$,  one cannot in general find explicit solutions to the Orr-Sommerfeld equation (\ref{OS}). In that case the wave resistance has to be evaluated numerically. For further details on approximation methods for the Orr-Sommerfeld equation, readers may refer to \cite{Kirby, helena}. In what follows we will thus assume the shear current $U(z)$ to be linear:  $U(z)=U_0(h+z)/h$, corresponding to a constant vorticity $\varpi = - U_0/h$
\cite{Vanden}.
This might be of relevance for tidal waves  \cite{Valenzuela, Pere} and  
constitutes a first step towards more general cases.
In that case $U''(z)=0$ and the Orr-Sommerfeld  equation (\ref{OS}) admits an exact solution
 \begin{eqnarray}
f_k(z)=(V-U_0)\frac{\sinh(k(z+h))}{\sinh(kh)}\label{solOS}.
\end{eqnarray}
Inserting (\ref{solOS}) into (\ref{WR}), the wave resistance is then given by 
\begin{eqnarray}
R=\lim_{\epsilon\rightarrow0}  \int \frac{dk}{2\pi\rho}\, \frac{-ik  | \hat P_{ext}(k)|^2 }{...},
\label{aveceps}
\end{eqnarray}
where the denominator (...) corresponds to
\begin{eqnarray}
 \left((U_0-V)^2-\frac{2i(U_0-V)\epsilon}{k}\right)k \coth(kh)  \nonumber \\
   -U'_0 \left(U_0-V-\frac{i\epsilon}{k}\right)- \left( g +\frac{\gamma k^2}{\rho} \right).
\end{eqnarray}

For a well localized pressure distribution of the form $p(x)=p \,\delta(x)$, one has $\hat P_{ext}(k)=p$. The corresponding wave resistance is shown in Fig.2 as a function of the reduced velocity $U_0/c_{\rm{min}}$ assuming $V=0$. To gain physical insight in the problem we also display in Fig.2 the wave resistance for a uniform current $U(z)=U_0$, both for finite and  infinite depth. We first observe, in the case of a uniform current with finite depth (black dashed curve), the existence of a shoulder at $U_0=\sqrt{gh}$. 
 This is related to the fact that when $k\rightarrow 0$, the phase velocity $\omega/k$ tends to a finite value $\sqrt{gh}$ (see insert in Fig.2) \cite{Lamb}. Hence, when $U_0$ becomes larger than $\sqrt{gh}$, one of the two poles in the integrand of equation (\ref{aveceps}) disappears. 
At first sight it is therefore surprising that no such shoulder is present in the case of linear current $U(z)=U_0(h+z)/h$
(blue solid line). 
In fact, one can show that in the case of linear current and $V=0$ the integrand of (\ref{aveceps}) conserves its
two poles  as $U_0$  increases \cite{Vnonnulle}.
\smallskip

We may also emphasize the fact that for both finite depth cases with uniform and linear current the minimum phase velocity is shifted from its value 
$c_{\rm{min}}= (4 g \gamma  /\rho)^{1/4}$ 
defined in the introduction of this letter for infinite depth. Therefore, in Fig.2, the solid blue curve as well as the dashed black curve do not diverge exactly at abscissa $U_0/c_{\rm{min}}=1$ but for quantities slightly superior to it. Nevertheless, for $h$ larger than say $3\kappa^{-1}$, the minimum phase velocity for the uniform current and finite depth case becomes very close to $c_{\rm{min}}$. 

\begin{figure}[H]
\begin{center}
\includegraphics[width= 1 \columnwidth]{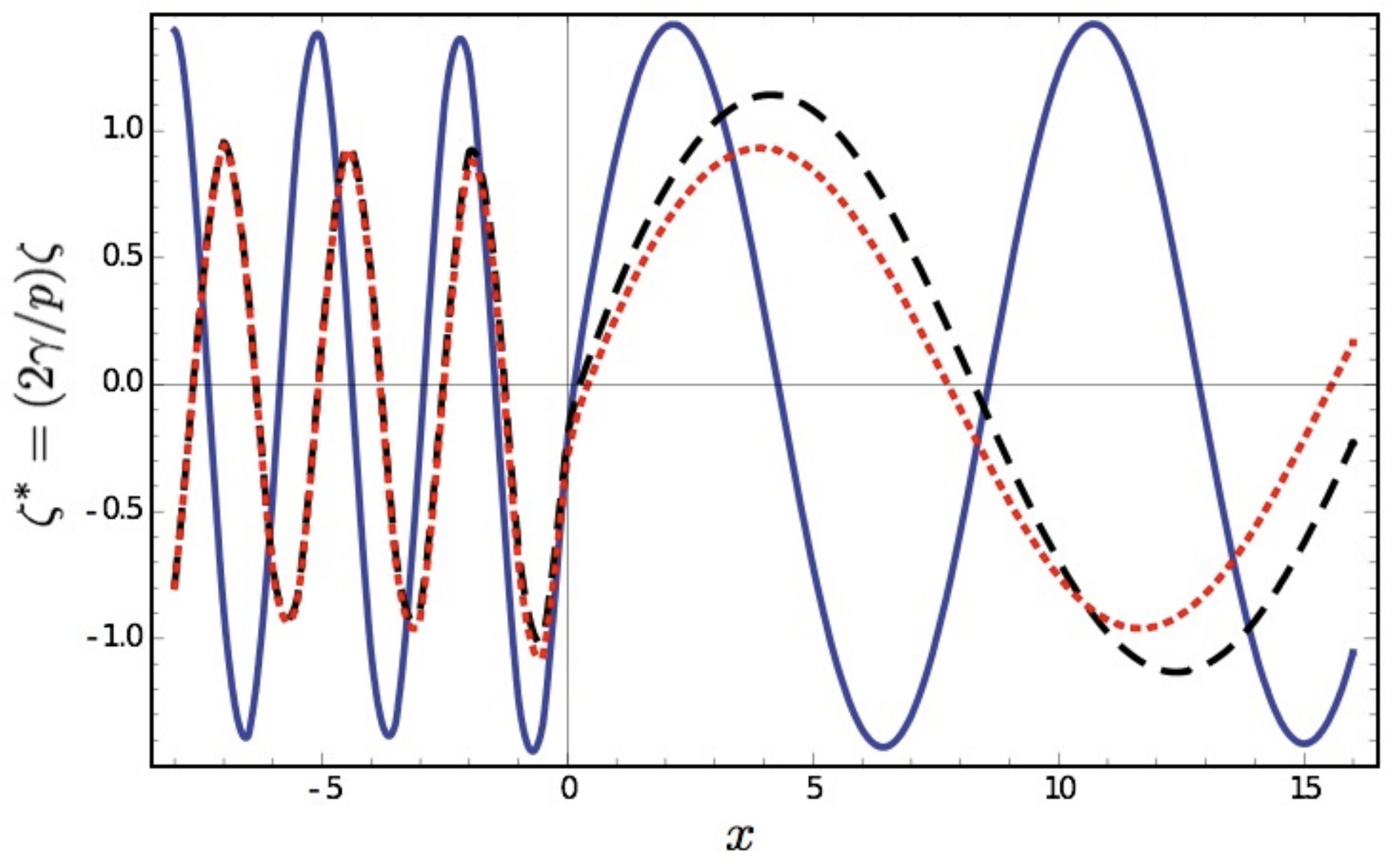}
\end{center}
\caption{(Color online) Plot of the free surface profile $\zeta(x,t)$ in units
of $p /2\gamma$, as a function space $x$ for $U_0/c_{min}=1.2$ and  $h=5 \kappa^{-1}$.  The red curve
(dotted) corresponds to a uniform current  with infinite depth, the black one (dashed) corresponds  also to a uniform current with finite depth $h$, while the blue one (solid line)  corresponds to a linear current $U(z)=U_0(h+z)/h$ with  finite depth $h$.} 
\label{fig3}
\end{figure}

Fore completeness, we have displayed in Fig.\ref{fig3} the elevation of the free surface in the case of a linear current as obtained  from (\ref{versPGG}) (assuming a well localized pressure distribution of the form $p(x)=p \,\delta(x)$) and compared it with the 
surface elevation for uniform currents (finite and infinite depth). 
 %%%%%%%%%%%%%%%%%%%%%%%%%%%%%%%%%%%%%%%%% 
 \smallskip
 
{\it{Conclusion - }}
We have shown  that non uniform currents have important effects on capillary-gravity waves generated
by a two-dimensional perturbation. Both the waves properties and the corresponding wave resistance are 
significantly modified when a current exists in the fluid. It would be of great interest to extend the results of the present letter to three-dimensional 
perturbation. A deeper understanding of the physical response of the wave system near $c_{\rm{min}}$ will also
require the introduction of nonlinear effects \cite{Dias, Parau, Grimshaw}.
\smallskip

\acknowledgments

We are grateful to Fr\'ed\'eric Chevy for fruitful discussions.


\begin{thebibliography}{0}
\bibitem{Vnonnulle} This is no longer true if $V\neq 0$. If the pressure distribution does move against the current, there is a critical velocity (that depends on depth $h$) for which the shoulder appears again.
\end{thebibliography}
\end{document}